\documentclass[10pt,conference]{IEEEtran} 
\IEEEoverridecommandlockouts
\usepackage{cite}
\usepackage{amsmath,amssymb,amsfonts}
\usepackage{algorithmic}
\usepackage{graphicx}
\usepackage{textcomp}
\usepackage{xcolor}
\def\BibTeX{{\rm B\kern-.05em{\sc i\kern-.025em b}\kern-.08em
    T\kern-.1667em\lower.7ex\hbox{E}\kern-.125emX}}
\usepackage{amsmath,bm}
\usepackage{comment}
\usepackage{enumitem}
\usepackage{siunitx} 
\usepackage{multirow}
\usepackage{booktabs}
\usepackage{soul}
\usepackage{pifont}
\usepackage{float} 
\usepackage{subfigure}

\usepackage[bookmarks=true,breaklinks=true,colorlinks,linkcolor=blue,citecolor=blue,urlcolor=blue]{hyperref}
\usepackage{xspace}

\newcommand{\design}{{CryptoQFL}\xspace}
\newcommand{\homo}{homomorphic}

\begin{document}

\IEEEoverridecommandlockouts

\title{\design: Quantum Federated Learning \\on Encrypted Data
}

\author{ 
    Cheng Chu$^1$, Lei Jiang$^{1,2}$, Fan Chen$^{1,2}$\\
	\textit{$^1$Luddy School of Informatics, Computing, and Engineering, Indiana University, Bloomington, IN}\\
    \textit{$^2$Quantum Science and Engineering Center, Indiana University, Bloomington, IN} \\
	E-mail: \{chu6, jiang60, fc7\}@iu.edu\\
 }
\maketitle

\begin{abstract}
Recent advancements in Quantum Neural Networks (QNNs) have demonstrated theoretical and experimental performance superior to their classical counterparts in a wide range of applications. However, existing centralized QNNs cannot solve many real-world problems because collecting large amounts of training data to a common public site is time-consuming and, more importantly, violates data privacy. Federated Learning (FL) is an emerging distributed machine learning framework that allows collaborative model training on decentralized data residing on multiple devices without breaching data privacy.
Some initial attempts at Quantum Federated Learning (QFL) either only focus on improving the QFL performance or rely on a trusted quantum server that fails to preserve data privacy.
In this work, we propose~\design, a QFL framework that allows distributed QNN training on encrypted data. 
\design is
(1) \textit{\underline{secure}}, because it allows each edge to train a QNN with local private data, and encrypt its updates using quantum \homo~encryption before sending them to the central quantum server;
(2) \textit{\underline{communication-efficient}}, as~\design quantize local gradient updates to ternary values, and only communicate non-zero values to the server for aggregation;
and (3) \textit{\underline{computation-efficient}}, as \design presents an efficient quantum aggregation circuit with significantly reduced latency compared to state-of-the-art approaches.
\end{abstract}



\section{Introduction}
\label{sec:introduction}
Recent advances in Quantum Neural Networks (QNNs)~\cite{tacchino2019artificial, perez2020data, cong2019quantum} using Variational Quantum Circuits (VQCs)~\cite{schuld2020circuit} have shown exponential quantum supremacy against classical neural networks in various quantum~\cite{huang2022qadv} and classical applications~\cite{havlivcek2019supervised}  
on today's noisy intermediate-scale quantum (NISQ) devices~\cite{preskill2018quantum}.
However, the training of QNNs relies on large amounts of training data that may be generated and hosted by multiple organizations. Integrating data to a common site by transporting the data across organizations is usually impossible in real-world situations due to data privacy, government regulations, or national security~\cite{savage2020results}.
To address this challenge, Federated learning (FL)~\cite{konevcny2016federated, yang2019federated} was proposed to decouple the model training from the need for direct access to the raw training data. However, FL is vulnerable to potential data leakage\cite{melis2019exploiting}. A malicious semi-honest server can leverage the uploaded local gradients to infer private data.

Inspired by the recent research on quantum homomorphic encryption~\cite{broadbent2015quantum, dulek2016quantum, mahadev2020classical}, we set out to address the aforementioned security challenges by presenting a privacy-preserving Quantum Federate Learning framework, referred as~\design, that allows distributed QNN training on encrypted data. 
This work makes the following contributions.

\begin{itemize}[leftmargin=*, topsep=0pt, partopsep=0pt]
\item \textbf We propose a baseline design for secure QFL using quantum homomorphic encryption, and evaluate its performance through experiments. Our experiments reveal the performance bottlenecks in the baseline design, which motivate the optimizations we introduce in our proposed~\design framework.

\item \textbf We propose the~\design framework, which features three key optimizations: (1) an optimized workflow that streamlines the QFL process, (2) the use of ternary gradients to reduce communication overhead, and (3) an efficient quantum adder circuit that significantly reduces the overall latency. Together, these optimizations lead to improved QFL performance in terms of both speed and accuracy.

\item \textbf We conduct comprehensive experiments to evaluate the performance of the proposed~\design framework using various quantum applications. Specifically, we demonstrate its improved performance in terms of speed and accuracy, as well as its scalability and convergence. These experiments provide empirical evidence that the~\design framework is a promising solution for quantum federated learning, especially for scenarios with large-scale distributed data and privacy concerns.
\end{itemize}

\section{Preliminary}
\label{sec:background}
\subsection{Threat Model}
In our threat model, we consider semi-honest corruptions~\cite{zhang2022federated,Araki:CCS2016} in a horizontal quantum FL setting. We assume that all parties, including a server and multiple clients, follow the protocol's description in software and hardware but attempt to infer information about the other party's input from the protocol transcript. Semi-honest adversaries act perfectly normal in terms of their public behaviors, making it difficult to detect their misbehavior. For example, a semi-honest server~\cite{Hatamizadeh_2022_CVPR} may reconstruct other parties' private training data by performing gradient-based inversion attacks. In this work, we propose \design to prevent untrusted quantum central servers from performing such attacks and inferring more information about the quantum gradients from clients, which is a practical threat model compared to prior works.

\begin{figure}[t!]\centering
\includegraphics[width=3.4in]{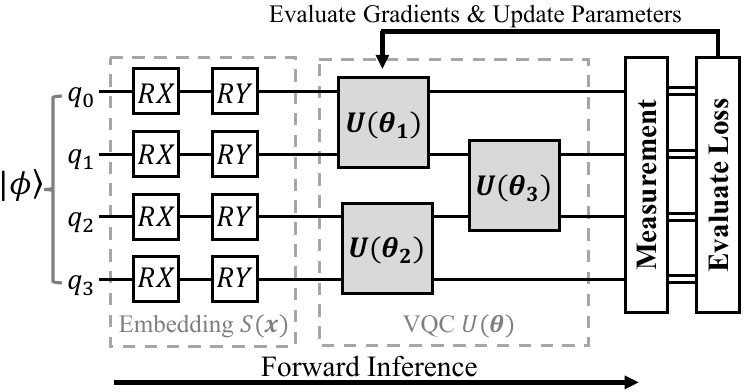}
\caption{A standard QML model.}
\label{f:background_qnn}
\end{figure}

\subsection{Quantum Computing} 
A quantum computing system leverages superposition of basis states to represent a $2^n$-dimensional complex Hilbert space $\mathcal{H}$=${({\mathbb{C}}^2)}^{{\otimes}n}$ with only $n$ quantum bits (qubits). The quantum state of a $n$-qubit system is described by a normalized vector $|\phi\rangle$=$\sum_{i=0}^{2^n-1}\alpha_i{|b_i\rangle}$, where $|b_i\rangle$ is the standard basis and $\sum_{i=0}^{2^n-1}|\alpha_i|^2$=$1$. Quantum measurements on a standard basis produce probabilistic outcomes that obey the Born rule: the probability for observing a measured result $|b_i\rangle$ is $|\alpha_i|^2$. 
A quantum gate operating on a $n$-qubit state multiplies a unitary $2^n$$\times$$2^n$ matrix \texttt{U} to a input state $|\phi\rangle$, resulting $|\phi^{'}\rangle$=\texttt{U}$|\phi\rangle$. One-qubit gates \texttt{X}, \texttt{Z}, \texttt{P}, \texttt{H} and two-qubit gate \texttt{CX} (i.e., controlled-X gate) generate the Clifford group, which can be seen as an analog to classical linear circuits that performs only additions. Adding any non-Clifford gates such as \texttt{T} gate or \texttt{CCX} gate (i.e., controlled-controlled-X gate, also known as Toffoli gate) to the Clifford group forms a gate set that is capable of universal quantum computations.

\textbf{Quantum Neural Networks}. 
Figure~\ref{f:background_qnn} illustrates a standard QML circuit, comprising a classical-to-quantum embedding layer (\texttt{S}$(\mathbf{x})$) that maps classical inputs ($\mathbf{x}$) into the quantum Hilbert space, followed by a trainable variational quantum circuit (VQC) (\texttt{U}$(\theta)$) that generates a predicted output via forward inference. The output is obtained via quantum state measurement and used to evaluate a predefined loss function. Note that an encoder is unnecessary when dealing with quantum datasets. This type of parametrized and data-dependent quantum computing system can be implemented on noisy intermediate-scale quantum (NISQ) devices and effectively trained using classical gradient descent~\cite{rumelhart1986learning} or its quantum variant~\cite{kerenidis2020quantum, sweke2020stochastic}.

\textbf{Quantum Aggregation}.
A quantum adder is a fundamental component in many quantum computing applications, including quantum federated learning (QFL). However, applying current full adders~\cite{sarma2018quantum, kumar2017optimal} directly to QFL gradient aggregation has certain limitations. Firstly, these adders require a certain number of ancillary qubits, leading to the production of garbage outputs that cannot be reversibly removed or may not be used later. This waste of resources becomes particularly problematic when implemented on NISQ quantum devices with limited qubits. Secondly, current carry propagates adders are designed to handle operands with many bits, involving complex logical designs and requiring additional quantum gates to operate. However, in QFL, complex logic is not necessarily needed, as the operands (i.e., gradients) can be quantized to 2 or 3 bits without significantly impacting model accuracy (as discussed in Section~\ref{sec:design}). 
\textit{Our preliminary study has identified these limitations, emphasizing the need for a more efficient quantum adder design for QFL}.

\begin{figure}[t!]\centering
\includegraphics[width=3.4in]{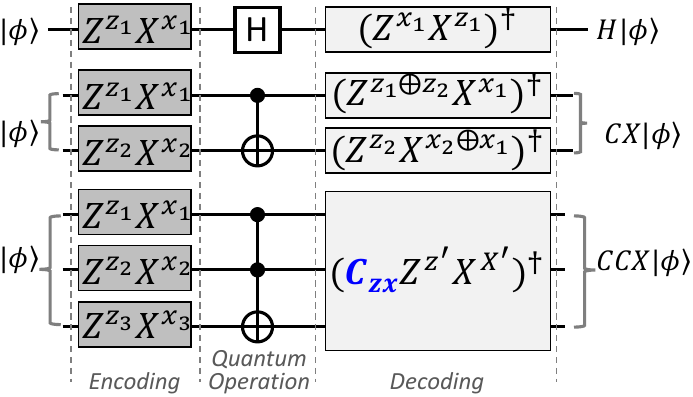}
\caption{QOTP keys update rule.}
\label{f:background_qhe}
\end{figure}

\subsection{Quantum Homomorphic Encryption} 
Homomorphic encryption (HE) allows computation on encrypted data to be performed by a party having access only to the ciphertext. Quantum homomorphic encryption (QHE) is the quantum analogue of classical HE, which enables the evaluation of quantum circuits on encrypted quantum data. Different state-of-the-art QHE schemes consider different non-Clifford gates, such as the \texttt{T} gate in~\cite{broadbent2015quantum, dulek2016quantum} and the \texttt{CCX} gate in~\cite{mahadev2020classical}. Despite the differences, they are all hybrid schemes that combine quantum one-time pad (QOTP) and CHE. The QHE computation consists of two parts: (1) QOTP encryption of the plaintext, and (2) CHE computation on the QOTP keys. Arbitrary quantum computation can then be applied directly to the encrypted quantum state. The homomorphic property of CHE is used to update the QOTP keys. Finally, QHE decryption can be performed using the encrypted results and the updated QOTP keys. 

\textbf{Quantum One-Time Pad}. 
QOTP encrypts an $n$-qubit state $|\phi\rangle$ with $n$ pairs of random binary
classical keys ($z_i$, $x_i$), where ${z_i, x_i}$$\in$\{0,1\} and $i$$\in$[1, n],
producing a maximally mixed state $|\phi_e\rangle$=$Z^{z}X^{x}|\phi\rangle$=$Z^{z_n}X^{x_n}$$\otimes$$\cdots$$\otimes$$Z^{z_1}X^{x_1}$$|\phi\rangle$
that is completely independent of the original state~\cite{ambainis2000private}.
To decrypt, the conjugate transpose (denoted as $\dagger$) of the original keys are 
applied on each qubit of $|\phi_e\rangle$,
producing $|\phi\rangle$=$(Z^{z}X^{x})^\dagger|\phi_e\rangle$.
QOTP provides a secure way to hide data rather than perform computations on it 
and has been widely used for quantum secure direct communication~\cite{deng2004secure, schumacher2006quantum}.

\textbf{QOTP Keys Update Rule in QHE}.
The \homo~application of a quantum gate \texttt{U} to a QOTP encrypted state can be represented as Equation~\ref{eq:gate_trans} and~\ref{eq:key_trans}.
We illustrate the QOTP key update rule for a \texttt{H} gate, a \texttt{CX} gate,
and a \texttt{CCX} gate in Figure~\ref{f:background_qhe}.
For a completed key update rules for all gates, we refer interested readers to~\cite{broadbent2015quantum, dulek2016quantum, mahadev2020classical}.
As it shows, if \texttt{U} is a Clifford gate, the updated QOTP keys ($z^{'}$, $x^{'}$) can be \homo{ally} computed following the \textit{Clifford scheme}~\cite{broadbent2015quantum}.
For instance,
the updated QOTP key for \texttt{H} and \texttt{CX} gates can be obtained through 
a swap or simple \texttt{XOR} operations.
The QOTP key update for non-Clifford \texttt{CCX} gates is more complicated.
As we highlighted in blue in Figure~\ref{f:background_qhe},
$C_{zx}$ consists of two \texttt{CX} gates and two \texttt{H} gates 
that is conditioned on the original QOTP keys ($z_i$, $x_i$).
One recent work~\cite{mahadev2020classical} constructed a scheme to perform such conditioned \texttt{CX} gates without knowing the plaintext of original QOTP keys,
but it involves additional quantum state preparation, measurement, and CHE computation~\cite{broadbent2015quantum, dulek2016quantum, mahadev2020classical},
resulting in significantly increased computing complexity and latency. 
\textit{Therefore, it is desired to reduce the number of \texttt{CCX} gates in a practical quantum circuits}.

\begin{equation}
{\mathbf{U}}(Z^{z}X^{x}|\phi\rangle)=Z^{z^{'}}X^{x^{'}}{\mathbf{U}}|\phi\rangle
\label{eq:gate_trans}
\end{equation}
\begin{equation}
{\mathbf{CHE}}(z, x) \rightarrow {\mathbf{CHE}}(z^{'}, x^{'}) 
\label{eq:key_trans}
\end{equation}

\begin{table}[t!]
\centering
\caption{Related work on quantum federated learning.}
\label{tab:related_work}
\begin{tabular}{|l|c|c|c|}\toprule
\multirow{2}{*}{\textbf{Scheme}} 
& \multirow{2}{*}{\textbf{Security}} 
&  \textbf{Communication} 
&  \textbf{Computation} \\
&  &  \textbf{Efficiency} &  \textbf{Efficiency} \\\hline\midrule
\cite{chen2021federated, xia2021quantumfed, huang2022qfl} 
& \ding{56} & \ding{52} & \ding{52} \\\hline
\cite{li2021quantum, sheng2017distributed, zhang2022federated}
& \ding{56} & \ding{52} & \ding{52} \\\hline
\text{Baseline}
& \ding{52} & \ding{56} & \ding{56} \\\hline
\textbf{\design}
& \ding{52} & \ding{52} & \ding{52} \\\bottomrule
\end{tabular}
\end{table}

\begin{figure}[b!]\centering
\includegraphics[width=3.4in]{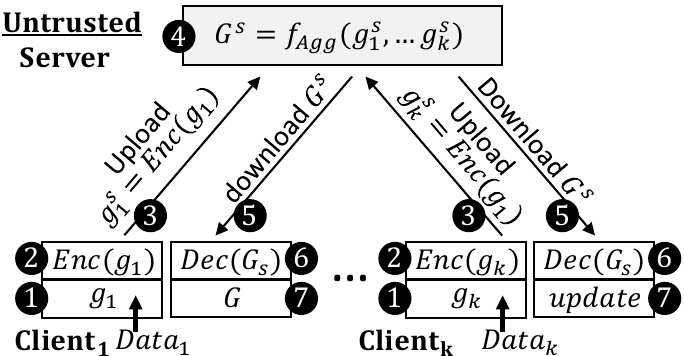}
\caption{Illustration of a general secure Federated Learning framework.}
\label{f:background_FL}
\end{figure}

\section{Related Work}
\label{sec:related}
Recent works in Quantum Federated Learning (QFL) \cite{chen2021federated, xia2021quantumfed, huang2022qfl, sheng2017distributed, zhang2022federated} have focused on adapting the FL framework to quantum machine learning. However, most efforts have been directed towards improving QFL model performance through various methods, including leveraging classical pre-trained models \cite{chen2021federated}, quantum fidelity-based loss functions \cite{xia2021quantumfed}, and new gradient optimization techniques \cite{huang2022qfl}. Unfortunately, few studies have considered the crucial issue of data privacy, with some only providing QNN inference without trainability \cite{li2021quantum} or assuming a trusted quantum aggregation server \cite{sheng2017distributed, zhang2022federated}. Moreover, previous research works have suffered from communication and computation inefficiencies, as summarized in Table~\ref{tab:related_work}.
Motivated by these limitations, we aim to develop a secure quantum federated learning framework that is both communication and computation efficient.

\section{Baseline Design and Analysis}
\label{sec:motivation}
Using the general secure federated learning framework~\cite{konevcny2016federated, yang2019federated} shown in Figure~\ref{f:background_FL}, we develop a secure QFL baseline. We present a detailed workflow of our approach and analyze its efficiency and time complexity.

\subsection{Working Procedure}
To set up the framework, each client is provided with a copy of the QNNcu model. The detailed working procedure is explained below.

\noindent
\textbf{Step{\large\ding{202}}: Quantum-classical hybrid training}.
Each client performs QNN training on a mini-batch of the private local data following the hybrid quantum-classical training method~\cite{sweke2020stochastic}.
The computed floating-point gradients $g$ is then embeded into the amplitude of a quantum state $|\phi_{g}\rangle$.

\noindent
\textbf{Step{\large\ding{203}}: QHE Encryption}.
We apply the two-step QHE encryption:
(1) the quantum gradient state $|\phi_{g}\rangle$ is QOTP encrypted using random key \{$x^j$, $z^j$\}, where $x$, $z$ are vectors, and $j$ denotes the cliend ID;
(2) the QOTP key \{$x$, $z$\} is CHE encrypted.

\noindent
\textbf{Step{\large\ding{204}}: Upload gradients qubits and QOTP keys to cloud}.
The QOTP encrypted gradient qubit $|\phi_{g}\rangle$ and CHE encoded QOTP key $x$, $z$ at each client are transmitted to the cloud aggregator.

\noindent
\textbf{Step{\large\ding{205}}: Gradient qubit aggregation and QOTP key update}.
All of the encrypted local gradient qubits are {\homo}ally aggregated using a baseline full adders~\cite{sarma2018quantum, kumar2017optimal}. The QOTP keys are updated following the update rule in QHE~\cite{broadbent2015quantum, dulek2016quantum, mahadev2020classical}.

\noindent
\textbf{Step{\large\ding{206}}: Downloading aggregated gradient and updated QOTP keys to the clients}.
The aggregated gradients $G^s$ and updated QOTP keys are dowloaded to local edges. 

\noindent
\textbf{Step{\large\ding{207}}: Decryption of Gradients}.
With updated QOTP keys, each client performs a CHE decryption to obtain the QOTP plaintext \{$x^{'}$, $z^{'}$\}, 
and then decrypt $G^s$ to obtain $G$.

\noindent
\textbf{Step{\large\ding{208}}: Model Update}.
Each client updates its local model using $G$.

\begin{figure}[t!]\centering
\includegraphics[width=3.4in]{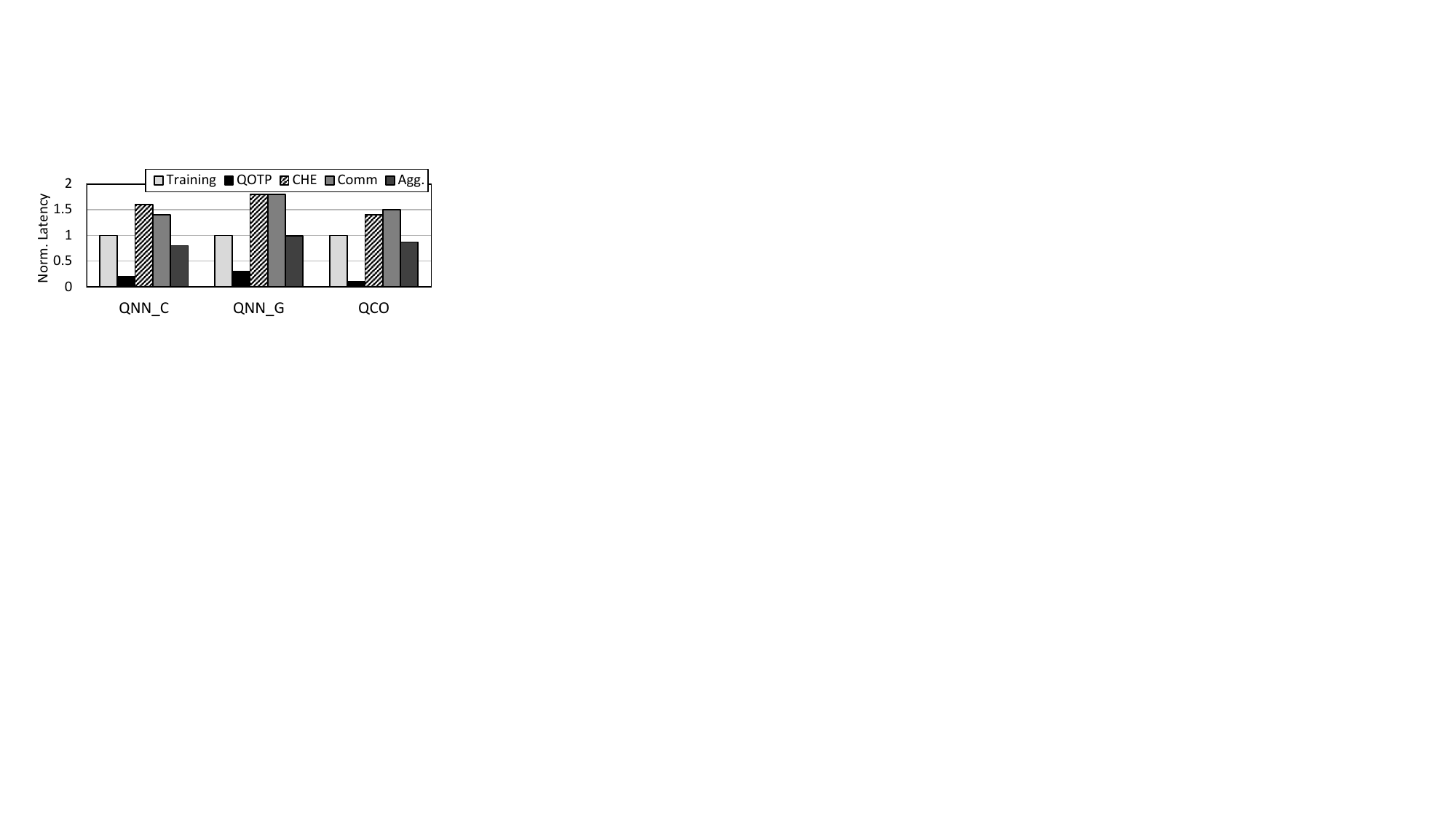}
\caption{Normalized latency breakdown in baseline design.}
\label{f:latency_base}
\end{figure}

\subsection{Efficiency and Time Cost Analysis}
We performed collaborative training on three different tasks using the baseline framework and reported the normalized latency breakdown in Figure~\ref{f:latency_base}. A detailed description of the experimental setup in this work can be found in Section~\ref{subsec:setup}.

It is evident that the CHE computation, communication of gradients, and aggregation incur significantly higher latency compared to the necessary latency required for local model training. These findings have motivated us to optimize the baseline design by considering the following factors:
\begin{enumerate}[leftmargin=*, topsep=0pt, partopsep=0pt, itemsep=0pt]
\item Optimizing the working procedure to reduce CHE computation time (Section~\ref{subsec:opt_flow}).
\item Quantizing the gradients to reduce communication costs (Section~\ref{subsec:terngrad}). 
\item Designing a more efficient quantum adder for QFL (Section~\ref{subsec:qadder}).
\end{enumerate}

\begin{table}[t]\centering
\caption{The accuracy of~\design on different tasks.}
\label{tab:terngrad_acc}
\begin{tabular}{|c|l|l|l|l|l|l|}
\toprule
\multirow{2}{*} {\textbf{Tasks}}  & \multicolumn{2}{c|}{\textbf{QNN\_C}}  & \multicolumn{2}{c|}{\textbf{QNN\_G}}   &  \multicolumn{2}{c|}{\textbf{QCO}} \\\cline{2-7}
    & Float  &  3-bit  & Float  &  3-bit & Float  &  3-bit \\\midrule
\textbf{Accuracy/}   & \multirow{2}{*} {99.25$\%$}  &  \multirow{2}{*} {99.15$\%$}  & \multirow{2}{*} {1}  &  \multirow{2}{*} {1} & \multirow{2}{*} {1}  &  \multirow{2}{*} {1}\\ 
\textbf{Fidelity}    &   &    &  &   &   &  \\  \bottomrule    
\end{tabular}
\end{table}

\begin{figure}[t!]\centering
\includegraphics[width=3.5in]{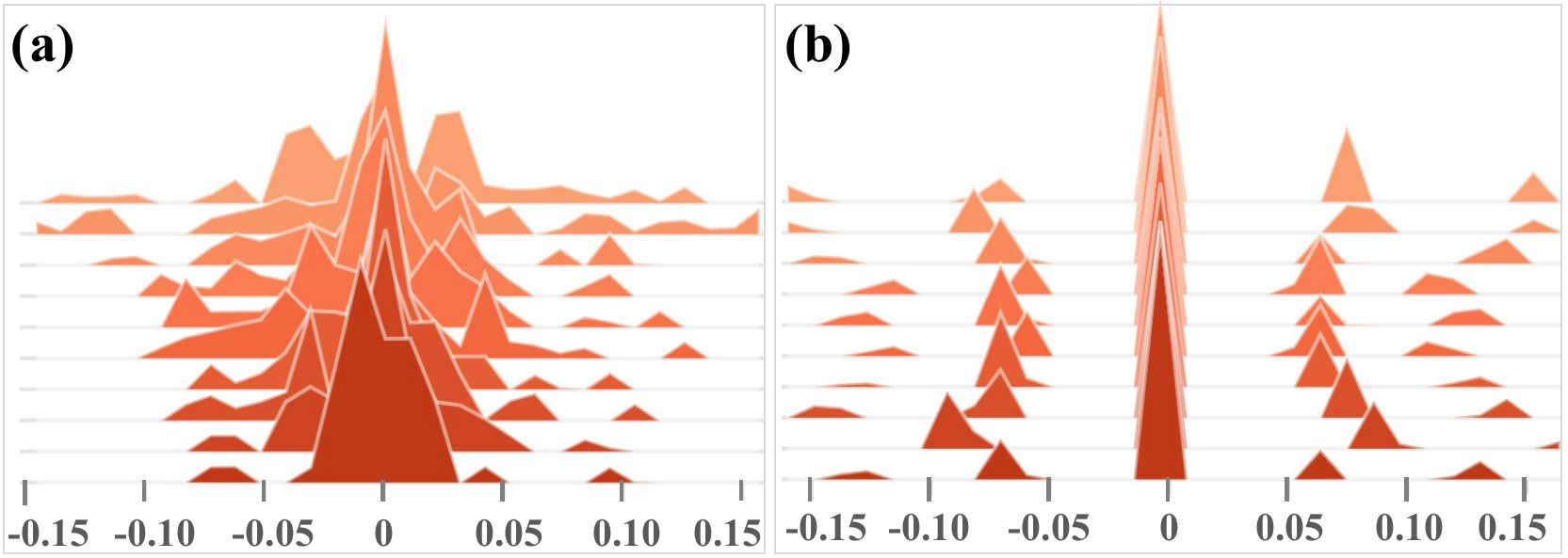}
\caption{TensorBoard visualized distribution of (a) floating gradients and (b) ternarized gradients for a supervised QNN classification model used in~\cite{QiskitQNN}.}
\label{f:tergrad}

\end{figure}

\section{\design}
\label{sec:design}
We propose, \design, a secure and efficient Quantum Federated Learning framework leveraging quantum \homo~encryption~\cite{broadbent2015quantum, dulek2016quantum, mahadev2020classical}. 
To reduce the overhead of CHE computing, we propose an updated QFL procedure that allows edges to share the same key without compromising accuracy or security. To address the communication inefficiency arising from computation on gradients with large bit widths, we propose using ternary quantization to reduce the bit width of QNN gradients. In addition, we optimize the baseline quantum full adders by designing a compact binary quantum adder for ternarized operands that only requires the use of the Clifford gate \texttt{CX} and the non-Clifford gate \texttt{CCX}.

\subsection{The \design Framework}
\label{subsec:opt_flow}
In our proposed~\design framework, we assume that clients are aware that the server is performing aggregation and have knowledge of the aggregation circuits used in the cloud. This assumption is well-supported and reasonable given the nature of federated learning, where clients participate in the learning process by contributing their local models to the cloud for aggregation.
With this approach, clients can use shared QOTP keys to encrypt their updated local model parameters (i.e., $g$), perform the CHE computation locally on their QOTP keys, and then only send the encrypted $g$ to the server for aggregation. This eliminates the need for the clients to send their QOTP keys to the server, which in turn reduces communication overhead and enhances the security of the QFL framework.
  
This approach not only improves the efficiency of the QFL framework but also enhances its security. By performing QOTP updates locally, clients can ensure that the QOTP keys are not exposed to the cloud or any other third party during the communication process. This also reduces the risk of potential security breaches or attacks on the communication channel. 
Accordingly, we highlight the optimized steps compared to the baseline procedure.

\noindent
\textbf{Optimized Step{\large\ding{203}}-\large\ding{206}}: 
The same random QOTP keys \{x, z\} are shared among all edges, which results in a reduction of the computation overhead by a factor of $N$, where $N$ is the total number of edges.
We only apply QOTP encryption to $|\phi_{g}\rangle$ and send it to the cloud for aggregation.
At the same time, edges  update the QOTP keys locally based on the sequence of the gates used in the quantum adder.

\subsection{Ternary Gradients to Reduce Communication}
\label{subsec:terngrad}

To reduce the size of gradient transfer in QFL, we propose an extension to the classical gradient quantization technique introduced in~\cite{wen2017terngrad}. Their approach uses ternary quantization, which represents the gradient values using only three levels: -1, 0, and 1. 
We build upon this approach and extend its application to QNNs.
Due to the cyclic nature of quantum parameters, where different quantum gates have specific modular scales, such as $2\pi$ or $4\pi$, we modified the baseline ternary scheme to a cyclical fashion. 
We conducted QNN training across different tasks, confirmed their convergence, and reported the corresponding accuracy/fidelity in Table~\ref{tab:terngrad_acc}. Additionally, we visually demonstrated the changes in gradient distribution of an example QNN model~\cite{QiskitQNN} in Figure~\ref{f:tergrad}.

\begin{figure}[t!]\centering
\includegraphics[width=3.4in]{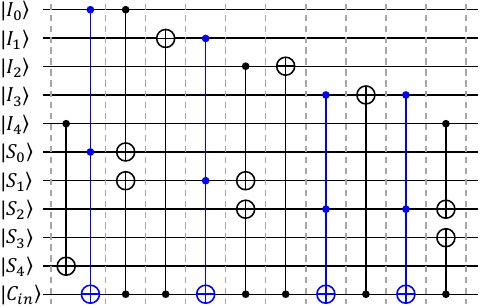}
\caption{The proposed quantum adder circuit.}
\label{f:adder}
\end{figure}

\subsection{A Fast Quantum Gradients Aggregator}
\label{subsec:qadder}
In the~\design framework,  gradient values are restricted to 0, 1, and -1. To facilitate the addition of signed binary numbers within this framework, we present a fast and efficient multi-bit quantum adder to reduce the overhead of aggregation computing.

\textbf{Proposed Design}.
To address the resource waste and hardware overhead associated with using previous quantum full adders \cite{sarma2018quantum, kumar2017optimal} for ternary value aggregation, we propose a new quantum adder circuit, as shown in Figure~\ref{f:adder}.
The quantum adder takes into account the input bits A, B, and the carry-in bit, resulting in eight possible combinations: 000, 001, 010, 011, 100, 101, 110, and 111. Among these combinations, 011, 101, 110, and 111 generate a carry-out. 
One direct approach to counting the carry-out is to use the Toffoli (i.e., \texttt{CCX}) gate. However, this method requires three Toffoli gates to compute each carry-out, resulting in significant cost and delay. To overcome this challenge, we use the first bit as the sign bit and obtain its complement. By operating on the complement, we only need one Toffoli gate to complete the carry-out computation, resulting in significant cost savings. Additionally, only two \texttt{CNOT} gates are needed to implement the complement operation. The circuit has no garbage output, and only the Toffoli gate is used as a non-Clifford gate. The partial sum can be accumulated on the original qubits, allowing for the serial operation of additional tasks by resetting the input qubits.


\begin{table}[t]\centering
\caption{Comparison of quantum aggregation schemes.}
\label{tab:Adder}
\begin{tabular}{|c|c|c|c|c|c|c|}\toprule
\textbf{Scheme} &\textbf{C\_in} &\textbf{Qubits\#} &\textbf{\texttt{CX}\#}  &\textbf{\texttt{CCX}\#}  &\textbf{Cost} &\textbf{Latency} \\\hline\midrule
QA1~\cite{li2020efficient}	&No &11	&15 &7	&65 &55 \\\hline
QA2~\cite{wang2016improved}  &No &16	&25 &5 &50 &50 \\\hline
\textbf{Ours}							&Yes &11 &10 &4	&30 &28\\\bottomrule
\end{tabular}
\end{table}

\textbf{Design Cost and Qubit Reset}.
In quantum circuits, the cost and latency of multi-qubit gates differ significantly, as shown in the comparison method proposed in~\cite{orts2020review}. To minimize the delay and overhead caused by the primary source, which is the Toffoli gate, we have significantly reduced its usage. 
We compare the propose quantum adder with previous works~\cite{sarma2018quantum, kumar2017optimal} in Table~\ref{tab:Adder}.
The \texttt{reset} operation has significantly lower latency compared to the \texttt{CNOT} gate, as reported in~\cite{basilewitsch2021fundamental} and confirmed by~\cite{das2021adapt, ibmq}. Consequently, it has a negligible impact on the overall system performance.

\section{Experiments}
\label{sec:experiments}

\subsection{Experimental Setup}
\label{subsec:setup}

\textbf{Benchmarks}.
We evaluate the \design framework using various classical and quantum applications. These applications include a supervised QNN provided by Qiskit~\cite{QiskitQNN}, an unsupervised QNN model provided by Pennylane~\cite{perez2020data}, and a combinatorial optimization solver provided by Paddle~\cite{QFinance}. 
For each model, we have followed its original configuration and adopted the proposed \design framework for collaboratively federated training.
We have summarized the benchmark applications used in this work, along with the corresponding datasets and code links, in Table~\ref{tab:benchmark}.

\textbf{Simulation}. 
To build the circuits for these three tasks, we utilize the APIs provided by Qiskit, Pennylane, and Paddle, respectively. By using these APIs, we can easily integrate the quantum circuits into the PyTorch workflows. The quantum circuit parameters can be naturally incorporated into PyTorch classical architectures and trained jointly without any additional operations. We follow the settings provided by the corresponding libraries for the learning rate, batch size, optimizer, and weight decay.

\begin{table}[t]\centering
\caption{Summary on evaluated benchmark applications.}
\label{tab:benchmark}
\begin{tabular}{|c|l|l|l|}\toprule
\textbf{Tag}  & \textbf{Application} & \textbf{Provider}  & \textbf{Dataset}  \\ \hline\midrule
QNN\_C & Supervised QNN~\cite{QiskitQNN}   & Qiskit & MNIST  \\\hline
QNN\_G & Unsupervised QNN~\cite{perez2020data}    & Pennylane  & Quantum States\\\hline
QCO & Comb. Optimization~\cite{QFinance} & Paddle & Real Stock\\ \bottomrule  \end{tabular}
\end{table}

\begin{figure}[t!]\centering
\includegraphics[width=3.4in]{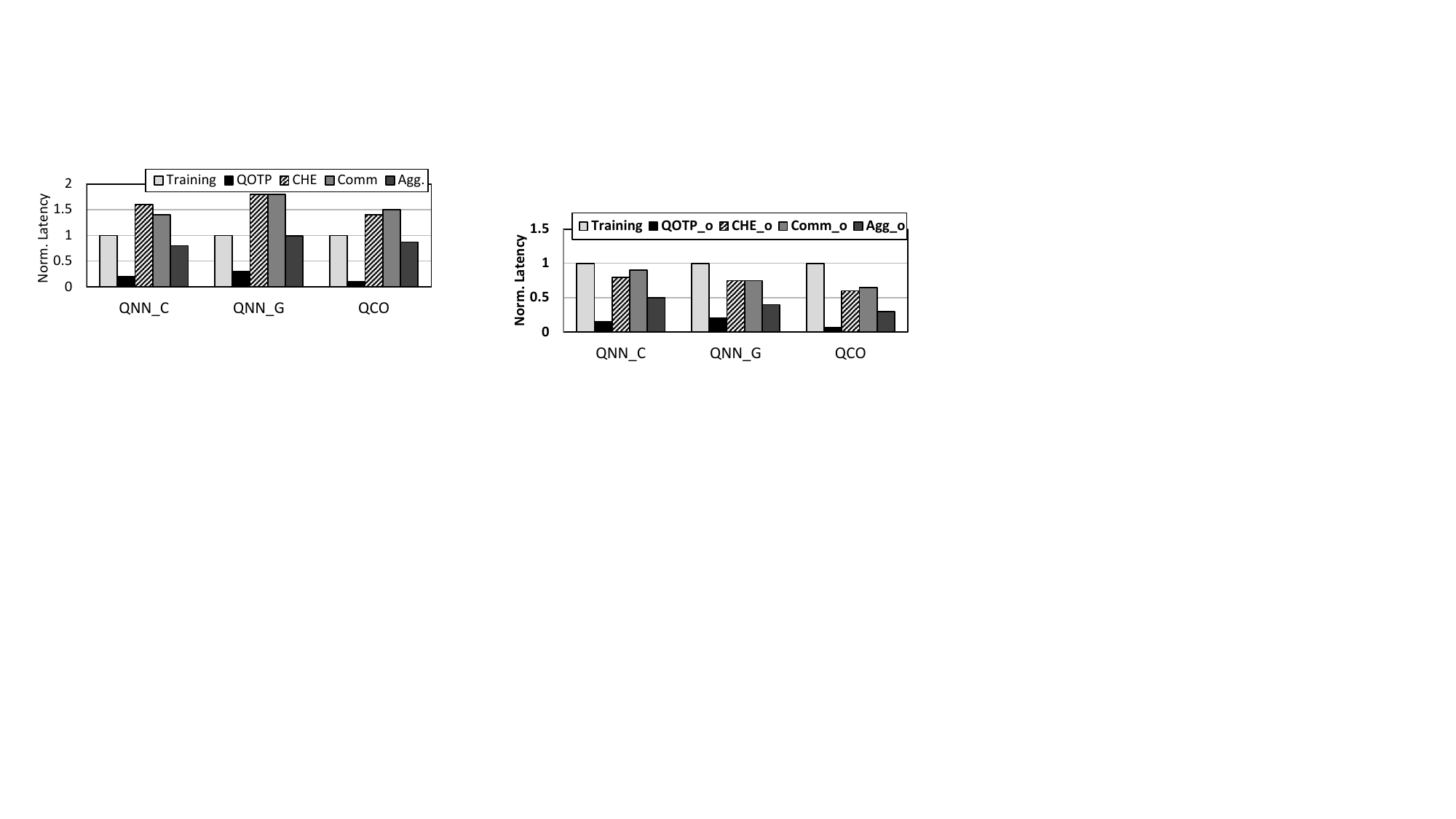}
\caption{Normalized latency breakdown in a~\design framework.}
\label{f:latency_opt}
\end{figure}
\begin{figure}[h!]\centering
\includegraphics[width=3.4in]{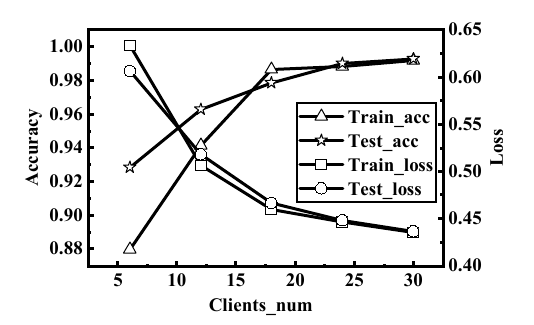}
\caption{QNN performance with scaling of client numbers using the QNN\_C~\cite{QiskitQNN} model on the MNIST dataset for 4-class classification.}
\label{f:client_num}
\end{figure}

\begin{figure*}\centering
\includegraphics[width=0.98\linewidth]{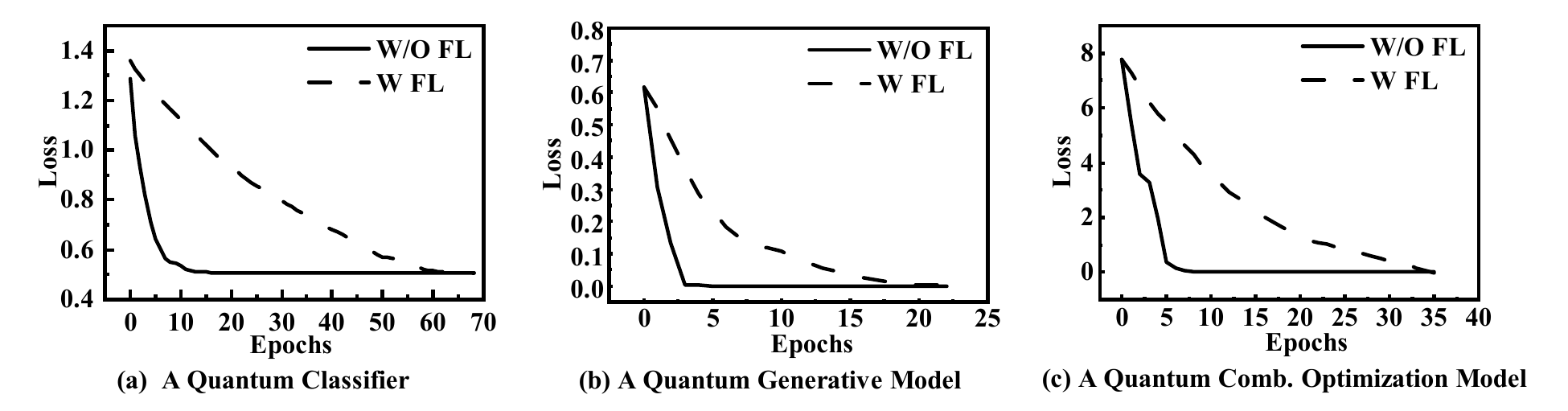}
\caption{Training loss v.s. training epochs on different tasks.}
\label{f:loss_iteration}
\end{figure*}

\subsection{Experimental Results and Analysis}
\label{subsec:results}

\textbf{Latency Analysis}.
We present the updated latency breakdown of the \design framework in Figure~\ref{f:latency_opt}. It shows the amount of time spent in each step of the federated learning process. Compared with the result for baseline design in Figure~\ref{f:latency_base}, the~\design scheme, which combines optimized working procedures, ternarized gradient, and a fast quantum adder, significantly reduces the latency bottleneck and improves the overall QFL performance. Specifically, the optimized design reduces the CHE computation time. The ternarized gradient reduces the communication time by reducing the number of bits required to transmit the gradient. Finally, the fast quantum adder improves the overall computation time of the quantum circuit. Together, these optimizations provide a significant reduction in the overall latency.

\textbf{Scalability with Clients Number}.
In Figure~\ref{f:client_num}, we compare the achieved final training loss and accuracy when the number of participating clients varies. We set each client with a fixed number of data samples, same as~\cite{chehimi2022quantum}. We observe that, in general, as the number of participating clients in the~\design setup increases, higher testing accuracy is achieved without overfitting the training data. 
The significant drop in accuracy for the case with only 5 clients is because the number of clients in a federated learning setup must be sufficiently large to achieve efficient learning.
However, we also observe that the QFL performance starts to plateau when the number of clients exceeds a certain threshold. This is because increasing the number of clients also increases the communication overhead and the amount of computation required for aggregation, which may offset the benefits of parallelism in the QFL approach.


\textbf{Convergence Analysis}.
To evaluate the convergence guarantee of the~\design framework, we perform numerical experiments in which all devices participate in the aggregation process. However, the convergence guarantee derived can be extended to cases where only a subset of devices are involved.
We compare the convergence speed of~\design with that of a non-federated training scheme in Figure~\ref{f:loss_iteration}. The results demonstrate that~\design not only converges but also achieves consistently similar performance in all three applications. This implies that the convergence of each local model to the global optimum is guaranteed in~\design.


\section{Conclusion}
In conclusion, recent advancements in Quantum Neural Networks have shown better performance than classical counterparts, but centralized QNNs face limitations due to data privacy concerns. Federated Learning is an emerging solution, but existing Quantum Federated Learning approaches have shortcomings. In this work, we proposed the~\design framework, which allows secure and efficient distributed QNN training using encrypted data. Our framework features three key optimizations that improve QFL performance in terms of speed and accuracy. Through comprehensive experiments, we demonstrated the scalability, convergence, and improved performance of~\design in various quantum applications. Our work provides empirical evidence that~\design is a promising solution for large-scale distributed data scenarios with privacy concerns.

\section*{Acknowledgments}
This work was supported in part by NSF CCF-1908992,
CCF-1909509, CCF-2105972, and NSF CAREER AWARD CNS-2143120.
Any opinions, findings, and conclusions or recommendations expressed in this material are those of the authors and do not necessarily reflect the views of grant agencies or their contractors.

\bibliographystyle{ieeetr}
\bibliography{main.bib}

\end{document}